\documentclass{article}
\usepackage{amsfonts,amsmath,amssymb}
\usepackage{graphicx}
\graphicspath{{figures/}}
\begin{document}

\title{Dynamics of Anisotropic Universes}

\author{J\'{e}r\^{o}me Perez \\ 
Laboratory of Applied Mathematics\\
 Ecole Nationale Sup\'{e}rieure de Techniques Avanc\'{e}es\\
  32 Bd Victor, 75739 Paris cedex 15 - France \\ jerome.perez@ensta.fr}
\maketitle

\begin{abstract}
We present a general study of the dynamical properties of Anisotropic Bianchi Universes in the context of Einstein General Relativity. Integrability results using Kovalevskaya exponents are reported and connected to general knowledge about Bianchi dynamics. Finally, dynamics toward singularity in Bianchi type {\sc viii} and {\sc ix} universes are showed to be equivalent in some precise sence.
\end{abstract}

\section{Homogeneous Universe and Bianchi models}

Considering the usual synchronous frame of General Relativity
\footnote{without contrary indications greek indexes run from 0 to 3, latin indexes from 1 to 3, metric signature is $(-,+,+,+)$,  $\varepsilon$ and $\delta$ are respectively the completely antisymetric Levi-Cevita tensor and the Kronecker symbol} $\ ds^{2}={g}_{\mu\nu}\,dx^{\mu}\,dx^{\nu}=\tilde{g}_{ij}\,dx^{i}\,dx^{j} -dt^{2}$. A Universe is said 
homogeneous when there exists an isometry group wich preserves the infinitesimal spacial lenght
$dl^{2}=\tilde{g}_{ij}\,dx^{i}\,dx^{j}$
A characterization of the isometry group is possible writing structure constants %
\begin{equation}
C_{ab}^{\;c}= \left( \partial _{i}e_{j}^{c}-\partial
_{j}e_{i}^{c}\right) e_{a}^{j}\,e_{b}^{i}  \label{construcgrou2}
\end{equation}
where $dx^{i}=e_{j}^{i}\,dy^{j}$. Constants $C_{ab}^{\;c}$ are tensorial, low components are antisymetrics, and follow the Jacobi rule : 
 \begin{equation}
C_{ab}^{\;e}\,C_{ec}^{\;d}+C_{bc}^{\;e}C_{ea}^{\;d}+C_{ca}^{\;e}C_{eb}^{%
\;d}=0 
 \end{equation}
Decomposing 
$C_{ab}^{\;c}=\varepsilon_{abd}\,N^{dc}+\delta_{b}^{c}\,A_{a}-\delta_{a}^{c}\,A_{b}\,$, where $N^{ab}$ is a symetric tensor,
one can show that equivalence classes of homogeneous universes {\sl are} equivalence classes of $N^{ab}$ with $N^{ab}A_{b}=0$. Without less of generality,  the symetry of $N^{ab}$ allows us to write
\begin{equation}
N^{ab}=\left[ 
\begin{array}{lll}
n_{1} & 0 & 0 \\ 
0 & n_{2} & 0 \\ 
0 & 0 & n_{3}
\end{array}
\right] 
\;\;\text{and} \;\;
A_{b}=\left[ a,0,0\right]  \label{nmatrix}
\end{equation}
Models split then into  Class A with $a=0$ and  Class B with $a\neq 0$
and can be arranged in the well known Bianchi models
\begin{center}
\begin{tabular}{c|ccc|c|c|}
& $n_{1}$ & $n_{2}$ & $n_{3}$ & $a$ & Model \\ \hline
$0$ is a triple eigenvalue of $N$ & $0$ & $0$ & $0$ & 0 &   $B_{\text{\textsc{I%
}}}$ \\  
& $0$ & $0$ & $0$ & $\forall $ & $B_{\text{\textsc{V}}}$ \\ \hline
$0$ is a double eigenvalue of $N$ & $1$ & $0$ & $0$ & 0 & $  B_{\text{\textsc{%
II}}}$ \\  
& $0$ & $1$ & $0$ & $\forall $ & $B_{\text{\textsc{IV}}}$ \\ \hline
$0$ is a simple eigenvalue of $N$ & $1$ & $1$ & $0$ & 0 & $  B_{\text{\textsc{%
VII}}_{o}} $ \\ 
& $0$ & $1$ & $1$ & $\forall $ & $B_{\text{\textsc{VII}}_{a}}$ \\ 
& $1$ & $-1$ & $0$ & $0$ & $  B_{\text{\textsc{VI}}_{o}}  $ \\ 
& $0$ & $1$ & $-1$ & $\neq 1$ & $B_{\text{\textsc{VI}}_{a}}$ \\ 
& $0$ & $1$ & $-1$ & $1$ & $B_{\text{\textsc{III}}}$ \\ \hline
$0$ is not an eigenvalue of $N$ & $1$ & $1$ & $1$ & $0$ & $  B_{\text{\textsc{%
IX}}}  $ \\ 
& $1$ & $1$ & $-1$ & $0$ & $  B_{\text{\textsc{VIII}}}  $%
\end{tabular}
\end{center}

\section{Einstein Equations}
Following \cite{macallum},  one writes 
$ds^{2}=\gamma \,\left( \tau \right) \omega
^{i}\omega ^{j}-N^{2}\left( \tau \right) d\tau ^{2}$ with 
$\gamma \left( \tau \right) 
=\mathrm{diag}\left[ e^{A_{1}(\tau)},e^{A_{2}(\tau)},e^{A_{3}(\tau)}\right]$ and $dt=N\left( \tau \right) d\tau$. The so called invariant differential forms basis $\omega ^{i}$ are linear combinations of $dx_{i}$ with exponential or trigonometric coefficient in $x_{i}$ (See \cite{macallum}). Finally, $\tau$ and  $N\left( \tau \right)$ are respectivelly the conformal time and the lapse function. 

\subsection{BKL Formalism}
This formalism was introduced in the 70's by \cite{BKL}.
Filling Universe by a barotropic fluid with pressure $P$ and energy density $\rho$ such that $P=\left( \Gamma -1\right) \rho $%
, taking $N^{2}\left( \tau \right)=V^{2}=e^{A_{1}+A_{2}+A_{3}}$ for the lapse function, some algebra then gives from Einstein equations, the equations for the dynamics of Bianchi Universes
\begin{equation}
\left\{ 
\begin{array}{rll}
0 & = &   E_{c} +  E_{p} +  E_{m} =H \\ 
\chi \left( 2-\Gamma \right) V^{2-\Gamma } & = & A_{1}^{\prime \prime
}+\left( n_{1}e^{A_{1}}\right) ^{2}-\left(
n_{2}e^{A_{2}}-n_{3}e^{A_{3}}\right) ^{2} \\ 
\chi \left( 2-\Gamma \right) V^{2-\Gamma } & = & A_{2}^{\prime \prime
}+\left( n_{2}e^{A_{2}}\right) ^{2}-\left(
n_{3}e^{A_{3}}-n_{1}e^{A_{1}}\right) ^{2} \\ 
\chi \left( 2-\Gamma \right) V^{2-\Gamma } & = & A_{3}^{\prime \prime
}+\left( n_{3}e^{A_{3}}\right) ^{2}-\left(
n_{1}e^{A_{1}}-n_{2}e^{A_{2}}\right) ^{2}
\end{array}
\right.   \label{dynbianchi}
\end{equation}
where $\chi =8\pi G c^{-4}$, 
$E_{c}=\frac{1}{2}\sum\limits_{i\neq j=1}^{3}A_{i}^{\prime }A_{j}^{\prime }$, $E_{p}=\sum\limits_{i\neq j=1}^{3}n_{i}n_{j}e^{A_{i}+A_{j}}
      -\sum\limits_{i=1}^{3}n_{i}^{2}e^{2A_{i}}$, 
$E_{m}=-4\chi \rho \;V^{2}$ and a $\;\prime$ stands for $d/d\tau$

\subsection{Hamiltonian formalism \label{secham}}
\begin{figure}
  \includegraphics[height=4.5cm]{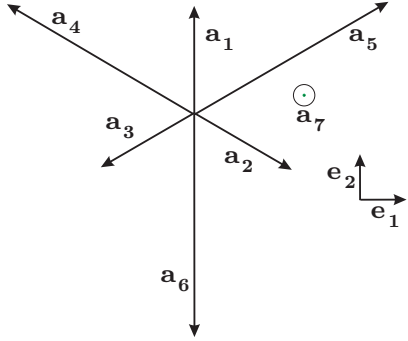}
  \caption{Projection of $\mathbf{a}_{i}$ vectors on $\left(  \mathbf{e}\text{\thinspace}_{1},\mathbf{e}_{2}\right)$ plane}
  \label{vectors}
\end{figure}

This formalism was introduced by \cite{misner1}, at almost the same time than the precedent. It consists to diagonalize the quadratic form $E_{c}$ by introducing new variables such that
 \begin{equation}
M:=\left[ 
\begin{array}{ccc}
\frac{1}{\sqrt{2}} & \frac{-1}{\sqrt{2}} & 0 \\ 
\frac{1}{\sqrt{6}} & \frac{1}{\sqrt{6}} & \frac{-2}{\sqrt{6}} \\ 
\frac{1}{\sqrt{6}} & \frac{1}{\sqrt{6}} & \frac{1}{\sqrt{6}}
\end{array}
\right] \;\;\;\;
\begin{array}{l}
\mathbf{q}:=\left[ q_{1}\;q_{2}\;q_{3}\right] ^{T}=M\;\left[
A_{1}\;A_{2}\;A_{3}\right] ^{T} \\ 
\\ 
\mathbf{p}:=\left[ p_{1}\;p_{2}\;p_{3}\right] ^{T}=M\;\left[ A_{1}^{\prime
}\;A_{2}^{\prime }\;A_{3}^{\prime }\right] ^{T}=\mathbf{q}^{\prime }
\end{array}
 \end{equation}
Dynamical equations for Bianchi Universes then becomes 
\begin{equation}
q_{1,2}^{\prime }=-\frac{\partial H}{\partial p_{1,2}} \; , \;\;\; 
p_{1,2}^{\prime }=-\frac{\partial H}{\partial q_{1,2}} \;\;\; 
\;\;\;\text{and}\;\;\; 
q_{3}^{\prime }= \frac{\partial H}{\partial p_{3}}  \; , \;\;\;
p_{3}^{\prime }=-\frac{\partial H}{\partial q_{3}}
\label{dynham}
\end{equation}
where $H=\frac{1}{2}\left\langle \mathbf{p},\mathbf{p}\right\rangle
+\displaystyle{\sum_{i=1}^{7}}k_{i}e^{\left( \mathbf{a}_{i},\mathbf{q}\right) }
$ with the following products
\begin{equation}
  \forall \mathbf{x},\mathbf{y}\in \mathbb{R}^{3}\;\;\;
        \left( \mathbf{x},\mathbf{y}\right) :=+x_{1}y_{1}+x_{2}y_{2}+x_{3}y_{3}  
        \;\;\;\text{and}\;\;\;
        \left\langle \mathbf{x},\mathbf{y}\right\rangle:=-x_{1}y_{1}-x_{2}y_{2}+x_{3}y_{3}
\end{equation}
and the  constants  $k_{1}:=2n_{1}n_{2}$,  $k_{2}:=2n_{1}n_{3}$,  $k_{3}:=2n_{2}n_{3}$, $k_{4}:=-n_{1}^{2}$,  $k_{5}:=-n_{2}^{2}$, $k_{6}:=-n_{3}^{2} $, $k_{7}=-4\rho _{o}\chi$. The set of vectors $\mathbf{a}_{i=1,\cdots,7}$ is highly symetric (see Fig. \ref{vectors}): It allows to use algebraic techniques based on Lie algebra in some dynamical treatments of this problem. 

\section{Integrability of Bianchi models}
\subsection{Kovalevskaya stuff and autosimilar ODE}

If an ODE 
\begin{equation}
\dot{\mathbf{x}}=\mathbf{f}\left( \mathbf{x}\right) \;\;\; \text{   with }\mathbf{x}\in \mathbb{R}^n \;\;\;\text{and}\;\;\; \dot{\ }\equiv d/dt
\label{nonlin}
\end{equation}
admits a self-similar solution 
\begin{equation}
\tilde{\mathbf{x}}=\left[   c_{1}  \left( t-t_{o}\right)
^{-  g_{1} },...,  c_{n} \left( t-t_{o}\right)^{- {g_{n}}}  \right] ^{T}
\end{equation}
with the weight vector $\mathbf{g}$ and the constant vector $\mathbf{c}$ lying respectively $ \mathbb{Z}^n$ and $ \mathbb{R}^n$. One can then prove that the linearized system around $\mathbf{\tilde{x}}$, admits too an auto-similar solution 
\begin{equation}
\mathbf{z}=
\left[ d_{1}\left( t-t_{o}\right) ^{  k_{1}- g_{1}},...,d_{n}
\left( t-t_{o}\right) ^{   k_{n} - g_{n}} \right] ^{T}
\end{equation}
where $\mathbf{d}$ is a constant vector, and, this time the Kovalevskaya vector $\mathbf{k}$, wich components are the so-called  Kovalevskaya exponents, lies $ \mathbb{C}^n$. In practice, one can compute this vector : it is easy to show that Kovalevskaya exponents are  eigenvalues of  $K:=D\left[\mathbf{f}\left( \mathbf{x}\right)\right] \left( \mathbf{ c }\right) +\mathrm{diag%
}\left( \mathbf{g}\right)
$.
A theorem by Poincar\'e shows that each component of the non linear general solution of equation (\ref{nonlin}) is on the form 
\begin{equation}
x_{i}\left( t\right)   \propto   \left( t-t_{o}\right) ^{-  g_{i}  }S\left[ \left(
t-t_{o}\right) ^{    k_{1} },...,\left( t-t_{o}\right) ^{    k_{n}  }\right] 
\end{equation}
where $S\left[.\right] $ stands for a multiple series. This result is at the basis of a sufficient condition (known as Yoshida's Theorem see \cite{Yoshida1},\cite{Yoshida2}) :  if all the Kovalevskaya exponents are  in $ \mathbb{Q}$, then the system is algebraicly integrable. 
\subsection{Kowalevskaya exponents and Bianchi Universes}
Such a work was Pioneering by \cite{Melnikov}, applied for the first time to some special cases by \cite{pavlov}, and more recently, developed with some imprecisions and incompletude by \cite{polonais}. We present here results obtained in a precise way in \cite{lape}.\\
Introducing new variables 
 \begin{equation}
\left\{ \mathbf{q},\mathbf{p}\right\} \mapsto \left\{ \mathbf{u},\mathbf{v}%
\right\} \text{ \ \ where \ \ }\left\{ 
\begin{array}{c}
\mathbf{u}\in  \mathbb{R}^{7}\text{, }u_{i=1,...,7}:=\left\langle \mathbf{a}_{i},%
\mathbf{p}\right\rangle  \\ 
\\ 
\mathbf{v}\in  \mathbb{R}^{7}\text{, }v_{i=1,...,7}:=\exp \left( \mathbf{a}_{i},%
\mathbf{q}\right) 
\end{array}
\right. 
 \end{equation}
Einstein's Hamiltonian equations become 
 \begin{equation}
\forall i=1,\cdots,7\;\;\left\{ 
\begin{array}{l}
v_{i}^{\prime }=u_{i}v_{i} \\ 
\\ 
u_{i}^{\prime }=\sum\limits_{j=1}^{7}W_{ij}v_{j}
\end{array}
\right. \;\;\text{with }W_{ij}:=-k_{j}\left\langle \mathbf{a}_{i},%
\mathbf{a}_{j}\right\rangle 
 \end{equation}
This dynamical system admits an autosimilar solution 
$\mathbf{\tilde{x}}=\left[ \lambda t^{  -1 },\mu
t^{  -2 }\right] ^{T}
$
where the constants $\left[ \mathbf{\lambda },\mathbf{\mu }\right]
\in  \mathbb{R}^{7}\times \mathbb{R}^{7}$ are solutions of the algebraic system 
\begin{equation}
\forall i=1,\cdots,7\;\;\left\{ 
\begin{array}{l}
\sum\limits_{j=1}^{7}W_{ij}\,\mu _{j}=-\lambda _{i} \\ 
\\ 
\lambda _{i}\,\mu _{i}=-2\mu _{i}
\end{array}
\right.   \label{sysalgeb}
\end{equation}

Taking into account that Rank$(W)=3$, it exists 45 distincts non trivial
solution for system (\ref{sysalgeb}), and then 45 sets of 14 Kovalevskaya exponents for all Bianchi Universes.
Analysis of such sets let us claim that : 
\begin{itemize}
\item In vacuum or with stiff matter ($\Gamma =2)$, excepted $B_{IX}$ and $%
B_{VIII}$, all other class A \thinspace Bianchi models \thinspace have
fractional Kovalevskaya exponents. 
\item With non stiff matter ($0\leq \Gamma <2)$ class A \thinspace Bianchi
models have at least one real or complex Kovalevskaya exponent, {\sl except} for 
$B_{\text{\textsc{I}}}$ with fractional values of $\Gamma $ and $B_{II}$ with $\Gamma$ fractional in $\left[0,\frac{11+\sqrt{73}}{3}\simeq0.82\right]$
\end{itemize}
These integrability indications are conforted by \cite{Maciejewski} for $B_{VIII}$ using Morales-Ruis Theory and by \cite{conte} using Painlev\'e's analysis for $B_{IX}$.

\section{Exact solutions}
  
\subsection{$B_{\text{\textsc{I}}}$ dynamics : The fundamental state}
      
 In vacuum, the general solution of  $B_I$ model could be explicited. The line element is
 \begin{equation}
       ds^2=t^{2p_1}dx_1^2+t^{2p_2}dx_2^2+t^{2p_3}dx_3^2-dt^2
 \end{equation}
where
\begin{equation}
\exists ! u\in[1,+\infty[\;\;\text{such that}\;\;
         \left \{ 
            \begin{array}{l}
               p_1=-\Omega u/(1+u+u^2) \\
               p_2= \Omega(1+u)/(1+u+u^2) \\
               p_3= \Omega u(1+u)/(1+u+u^2) 
            \end{array}
          \right.
\end{equation}
The real $\Omega$  is directly proportional to Universe Volume variation wich is constant in $B_I$. 
Toward singularity ($t\rightarrow 0$), $x_1$ expands when $x_2$ and  $x_3$ contracts, all with a constant exponential rate. Any couple $[u,\Omega]\in[1,+\infty[\times\mathbb{R}$ associated to an order of axis defines a Kasner state.

\subsection{ $B_{\text{\textsc{II}} }$ dynamics : One Kasner transition}
 As noted by \cite{BKL}, $B_{\text{\textsc{II}}}$ dynamics corresponds to a transition between 2 asymptotic Kasner states :
 \begin{equation} 
 \left.    
      \begin{array}[c]{c}
      \left[  u,\Omega \right] \;\; \left(  \Box \triangle \Diamond  \right) \\
      \begin{array}[c]{ccccc}
           p_{1}& < &  p_{2}    & < & p_{3}      \\
           \Box &   & \triangle &   & \Diamond    
      \end{array}          \\
      \text{Initial Kasner State : } t\rightarrow +\infty
      \end{array}
 \;\right]        
 \;\leadsto \;
 \left[ \;     
        \begin{array}[c]{c}%
        \left[  u-1,\Omega(1-2p_{1})\right]  \ 
        \left( \triangle \Diamond \Box     \right)\text{ if }u\geq2\\
              \\
           \left[  \left(  u-1\right)  ^{-1},\Omega(1-2p_{1})\right]  \ \left(
           \triangle \Box \Diamond \right)\text{ if }u<2\\
            \\
            \text{Final Kasner State : } t\rightarrow 0
         \end{array}
 \right .   
         \label{b2transi}   
 \end{equation}

\subsection{Conjectures (partially proven)}
As indicated by the rigorous proof by \cite{ringstrom}, all classes of Bianchi models converge generically toward a simple Kasner state or a closure of Kasner states when $t\rightarrow 0$. Numerical analysis, more precisely Billiard analogy (see next section), let us think that : 
\begin{itemize}
\item $B_{\text{\textsc{VI}}_{o}}$ and $B_{\text{\textsc{VII}}_{o}}$ dynamics correspond to  a finite number of Kasner transitions 

\item $B_{\text{\textsc{VIII}}}$ and $B_{\text{\textsc{IX}}}$ dynamics correspond to an infinite number of Kasner transitions
\end{itemize}
\section{Billiard analogy}
\begin{figure}
  \includegraphics[height=5.5cm]{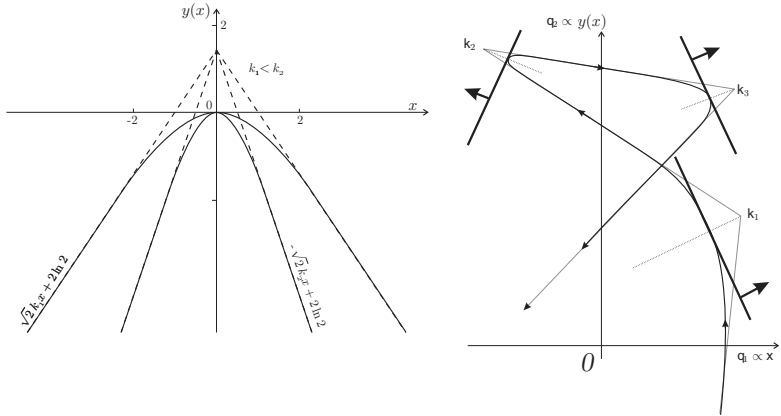}
  \caption{Reflexion on a straight cushion (left) and in a corner (right)}
  \label{rebonds}
\end{figure}

 Those works was pionneered by \cite{misner1}, and more recently by \cite{Uggla} or \cite{Jantzen}.
Introducing the super-time $\tilde{t}$ such that $d\tilde{t}=V^{1/3}dt$,
a "mass" $m=V^{4/3}$ and an "energy" $E=(dV/dt)^2 / 2V^{2/3}$, dynamical equation (\ref{dynham}) in vacuum becomes 
\begin{equation}
\left\{
\begin{array}[c]{c}%
\dfrac{dq_{1,2}}{d\tilde{t}}=\dfrac{p_{1,2}}{m}=\dfrac{\partial E}{\partial
q_{1,2}}\\
\\
\dfrac{dp_{1,2}}{d\tilde{t}}=-\dfrac{\partial\xi}{\partial q_{1,2}}%
=\dfrac{\partial E}{\partial p_{1,2}}%
\end{array}
\right.  \ \ 
\text{with}\ \
 \begin{array}[c]{l} 
    E=\dfrac{p_{1}^{2}+p_{2}^{2}}{2m}-\xi\left(  \mathbf{q} \right) \;, \;\;
    \xi\left(  \mathbf{q} \right)  = \sum\limits_{i=1}^{6}
    k_{i} e^{\left(
           \mathbf{\pi}\left(  \mathbf{a}_{i}\right)  ,\mathbf{q}\right)} \\   
           \mathbf{q}\in \mathbb{R}^2 \;
           ,\pi\;:\;\text{normal projector onto}\left(  \mathbf{e}\text{\thinspace}_{1},\mathbf{e}_{2}\right)\;\text{plane}            
  \end{array}
  \label{bil1}
\end{equation}
Toward singularity, $E\rightarrow+\infty$ and $m\rightarrow0$. When we go through time backward, equations (\ref{bil1}) are ones of a ball with a decreasing mass $m$ and with an increasing energy.
The exponential nature of the potential $\xi$ allows a precise description of he dynamics, namely, the billiard analogy.  When $\xi$
 is exponentially negligible, solutions of (\ref{bil1}) are almost straight lines in phase plane $(q_1 , q_2)$ : these are almost Kasner states in BKL representation. When $\xi$ cannot be neglected anymore, it is mainly due to one of it exponential term, equation of motion $-$ for a linear combination $y$ of $(q_1,q_2)$ $-$ takes a form equivalent to\footnote{The $x$ variable represents the supertime $\tilde{t}$ such that when $\tilde{t}\sim x$ one has $\xi \sim -k^{2}e^{y}$} 
\begin{equation}
\frac{d^{2}y}{dx^{2}}=-k^{2}e^{y}\ \ \ \ \ \text{with  }y\left(  0\right)
=\left.  \frac{dy}{dx}\right\vert _{x=0}=0
\label{analitik}
\end{equation}
This equation could be solved analytically in
\begin{equation}
y\left(  x\right)  =\ln\left[  1-\mathrm{th}^{2}\left(  \frac{kx}{\sqrt{2}%
}\right)  \right]  =-2\ln\left[  \mathrm{ch}\left(  \frac{kx}{\sqrt{2}%
}\right)  \right]
\end{equation}
It shows clearly a transition between the two Kasner states associated to the straight lines which are the asymptotes of this solution (see Fig. \ref{rebonds}). This generalized $B_{\text{\textsc{II}} }$ transition corresponds to a bounce of a ball in an amazing billiard. As a matter of fact, when $p_{1}^{2}+p_{2}^{2}=0$, all ball's energy is concentrated in the potential term $\xi$, this situation corresponds to the maximum of the $y$ function plotted in Fig. \ref{rebonds} : this is a bounce. It happens effectivelly against the isocontour line $\xi=\xi(\mathbf{q}^{*})$ where $d\mathbf{q}^{*}/d\tilde{t} = 0$. As we go toward singularity $(\tilde{t}\rightarrow 0)$ these lines move outward as one can see in Fig. \ref{contours}. The Universe ball then moves in a expanding billiard. It is easy to prove that ball's velocity norm $\mid d\mathbf{q}/d\tilde{t} \mid$ is always greater than the one of cushions which is the energy variation (see \cite{bd} for instance). This property remains true when barotropic matter fills Universe provided that $\Gamma <2$. The amplitude of the bounce depends on $k$ : the largest $k$ is, the smallest is the bounce angle between the two asymptotic linear regimes. Between the two ideal cases (quasi Kasner state and bounce) the parameter $k$ is dynamical, its variations remains negligible during each phase.  
\section{Dynamical properties of Bianchi Billiards}
\begin{figure}
  \includegraphics[height=9cm]{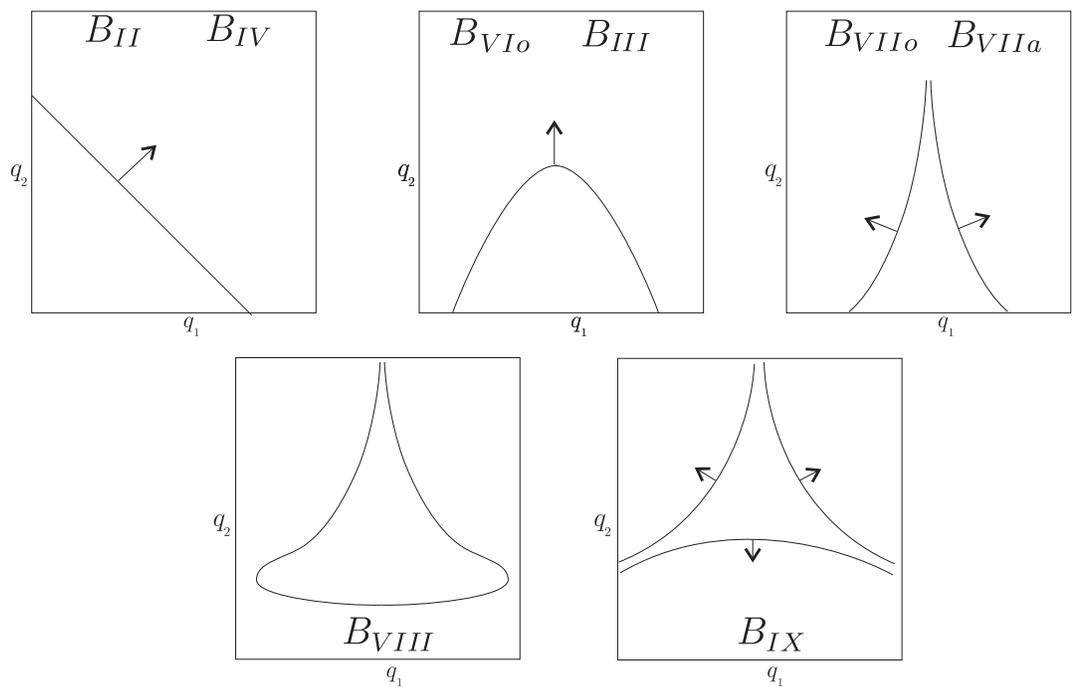}  
  \caption{Isocontours of potential $\xi$ for all distincts Bianchi Universe in vacuum. Arrows indicate the increasing values of $\xi$}
  \label{contours}
\end{figure}

In the billiard formulation, dynamics of Bianchi Universe is fully understood considering  isocontours of the potential $\xi$ which are presented on Fig. \ref{contours}.
Excluding $B_{\text{\textsc{IX}}}$ and $B_{\text{\textsc{VIII}}}$ all isocontours are open curves. Hence after a finite number of bounces, the ball representing Universe find a hole and go away to reach the asymtotic regime solution of equation (\ref{analitik}) which is a Kasner state. By opposite, for the two closed curves potential, there is no hole in the billiard, cushion allways expands less quickly than the ball, and there is an infinite sequence of $B_{\text{\textsc{II}}}$ transitions toward singularity. Although $B_{\text{\textsc{VIII}}}$ cushions seems open along a vertical channel, the bounce process makes the ball which enter the channel goes back. As sketched on Fig. \ref{rebonds}, in the channel $k\sim q_2$, then as the ball sinks into this region, $q_2$ grows and bounces are more and more pinched. By consequence, as the cushion is not strictly vertical, the ball finally goes back and leaves the channel.
These results are surely well known and one remaining question could be : Is  $B_{\text{\textsc{VIII}}}$ dynamics equivalent in some precise sence to $B_{\text{\textsc{IX}}}$ one ? In terms of complexity the two dynamics seems equivalent as they have exactly the same set of Kovalevskaya exponents. Let us show that they are also equivalent by the fact that they present the same kind of transition to chaos, and they have the same kind of attractor.
\subsection{$B_{\text{\textsc{IX}}}$ and $B_{\text{\textsc{VIII}}}$ Poincar\'e map analysis}
\begin{figure}[h]
\begin{center}
  \includegraphics[height=3.5cm]{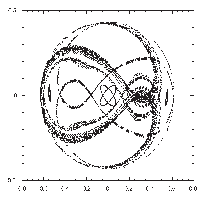}
  \includegraphics[height=3.5cm]{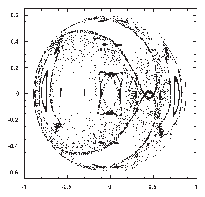}
  \includegraphics[height=3.5cm]{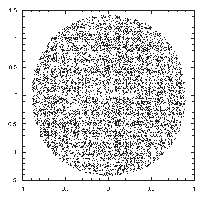} \\
  \includegraphics[height=3.5cm]{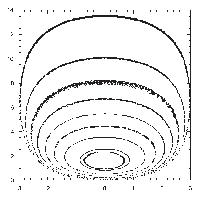}
  \includegraphics[height=3.5cm]{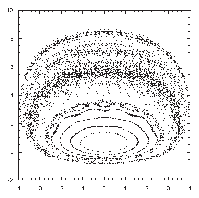}
  
  \caption{Poincar\'e maps toward singularity of temporal sections of $m=cst$ and $E=cst$ of  $B_{\text{\textsc{IX}}}$ (top : $m=1$ and $E=-2.8,-2.1,0$ from left to right) and $B_{\text{\textsc{VIII}}}$ (bottom : $m=1$ and $E=5.0,10.0$ from left to right) dynamics. For each map a set of 9 homogeneously distributed initial conditions have been choosen.}
  \label{poincare}
  \end{center}
\end{figure}

Detailled analysis show that Bianchi Universe volume is a regular variable. It is the only variable attached to cushions expansion in the billiard formalism. Therefore, the complexity of the dynamics is contained in the cushions's shape. In order to study this complexity in the two relevant cases which are $B_{\text{\textsc{IX}}}$ and $B_{\text{\textsc{VIII}}}$ Universes, we have studied dynamical propertied of temporal sections of the hamiltonian $E$ involving in equation (\ref{bil1}). As $V,\;t,\;\tilde{t}$ and $\tau$ are all related in a bijective way, by temporal section we mean sections of the dynamics described by equation (\ref{bil1}) with $m=cst$ and $E=cst$. For convenience we use always a ball with unit mass, we focuss attention under Poincar\'e maps associated to growing values of $E$. Such maps are obtained by considering the intersection between an orbit (a solution of the 4D differential system) and a fixed plane, namely $\Pi=\left\{ q_2=0,p_2=0 \right\}$ for maps plotted in Fig. \ref{poincare}. 
The greater are values of the energy $E$, the closest from singularity sections are. Analysis of such maps is clear using KAM Theory : When periodic or quasiperiodic orbits correspond to Poincar\'e sections disposed along curves, chaotic orbits sections fill dense regions. We can then conclude that far from singularity orbits are regular in both $B_{\text{\textsc{IX}}}$ and $B_{\text{\textsc{VIII}}}$. Approaching singularity, a transition to chaos seems to appear in the dynamics of both these closed cushions expanding billiards. In order to produce a comparative measure of this chaos we propose to produce an analysis base on $B_{\text{\textsc{IX}}}$ and $B_{\text{\textsc{VIII}}}$ Universes truncated dynamics.
\subsection{$B_{\text{\textsc{IX}}}$ and $B_{\text{\textsc{VIII}}}$ fractal attractors}
\begin{figure}
  \includegraphics[height=4.5cm]{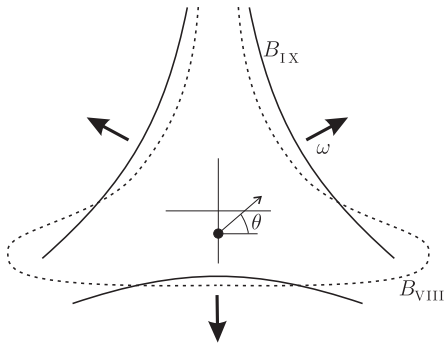} 
  \caption{Initial conditions $\theta$ and $\omega$ for fractals maps represented on Fig. \ref{fractals}}
  \label{condinit}
\end{figure}

\begin{figure}
  \includegraphics[height=5.5cm]{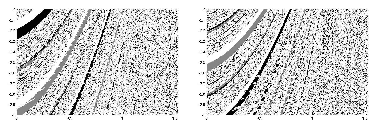} 
  \caption{$B_{\text{\textsc{VIII}}}$ (left) and  $B_{\text{\textsc{IX}}}$ (right) fractal maps in the $\left[  0,\pi/2\right]
\times\left[  -2,-3\right]  $ portion of the $\theta-\omega$ plane}
   \label{fractals}
\end{figure}

   As suggered by all previous indicators (Kovalevskaya exponents, Poincar\'e maps), we suspect $B_{\text{\textsc{VIII}}}$ dynamics to possess the same kind of attractor exhibited initially by \cite{cl} for $B_{\text{\textsc{IX}}}$. In order to confirm this suspition we have applied the truncated dynamical technique used by \cite{cl} in the context of Hamiltonian formalism described above.\\
   Initial conditions we used are $\mathbf{q}_o=\left[0,-\sqrt{6}\ln 2,\Omega\right]$ and $\mathbf{p}_o=\left[\cos\theta,\sin\theta,\omega\right]$ where the constraint $H=0$ is fullfilled provided that
 \begin{equation}
\Omega=\frac{3}{\sqrt{6}}\ln\left[  \pm
\frac{\left(  1-\omega^{2}\right)  }{2^{4}\left(  1\mp2\right)
}\right]
 \end{equation}
The upper and lower signs are associated respectively to $B_{\text{\textsc{IX}}}$ and $B_{\text{\textsc{VIII}}}$ Universes. The two free parameters $\omega$ and $\theta$ fixe respectively the initial variation rate of Univers volume (or the initial escape speed of cushions in the billiard formalism) and the orientation of the initial velocity vector. These initial conditions are represented on Fig. \ref{condinit}. 
The numerical integration of the dynamics from these initial conditions shows an unlimited sequence of $B_{\text{\textsc{II}}}$ transitions between Kasner states (see formula (\ref{b2transi})). We stop numerical integration when the Kasnerian $u$  parameter is greater than a critical value (for maps presented on Fig. \ref{fractals} we have used $u_{exit}=8$). When the dynamic is stopped, we affect  to the corresponding point in the $\theta-\omega$ plane a color which correspond to the expanding axe of stopped Kasner state (Red
$\leftrightarrow A_{1},$ Blue $\leftrightarrow A_{2}$ and Green
$\leftrightarrow A_{3}$). We have studied the $\left[  0,\pi/2\right]
\times\left[  -2,-3\right]  $ portion of the $\theta-\omega$ plane, projected on a regular $500\times500$ grid. The corresponding maps are presented in Fig. \ref{fractals}.
From this map we can compute the 3 Haussdorf dimensions associated to each color set. We can resume these three numbers into one by averaging each dimension weighted by the surface proportion occupied by the corresponding color. We then obtain a mean Haussdorf dimension of each map which is $d=1.6976\pm8.9\times10^{-3}$ for $B_{\text{\textsc{VIII}}}$'s map and $d=1.7141\pm8.5\times10^{-3}$ for $B_{\text{\textsc{IX}}}$'s map. These two numbers are then numerically indistinguable. In this sense we claim that $B_{\text{\textsc{VIII}}}$ and $B_{\text{\textsc{IX}}}$ universes are dynamically equivalent


\begin{thebibliography}{9}

\bibitem{macallum}M. Mac Callum, \emph{Anisotropic and Inhomogeneous Relativistic
Cosmologies}, in $\left[ 2 \right]$, pp 533--580   (1979)

\bibitem{HawkingIsrael}S.\ Hawking, W.\ Israel, \emph{General relativity :
an Einstein Centenary Survey}, Cambridge University Press  (1979)

\bibitem{BKL}V.\ A.\ Belinskii, L.M.\ Khalatnikov, E.M.\ Lifshitz,
\emph{Adv.\ Phys}, \textbf{19}, pp 525 (1970)

\bibitem{misner1}C.W.\ Misner, 
\emph{Phys.\ Rev.\ Let.}, \textbf{22}, pp 1071  (1969)

\bibitem{Yoshida1}H. Yoshida, \emph{Cel. Mech.}, \textbf{31},363 (1983)

\bibitem{Yoshida2}H. Yoshida, \emph{Cel. Mech.}, \textbf{31},381 (1983)

\bibitem{Melnikov}V.R.\ Gavrilov, V.D.\ Ivashchuk et V.N.\ Melnikov, 
        \emph{J. Math. Phys.}, \textbf{36},  5829 (1995) 
\bibitem{pavlov} V.I.\ Pavlov,  \emph{Regular and chaotic behaviour}, \textbf{1}, 111 (1996)

\bibitem{polonais}M.\ Szydlowski, M.\ Biesiada, 
 \emph{J. of Non Lin. Math. Physics}, \textbf{91},L1 2002

\bibitem{lape}J. Larena, J. Perez, \emph{Clas. \& Quant. Gravity}, in preparation
 
 \bibitem{conte}R.\ Conte,
pp 77-180, in \emph{The Painlevé property : One century later},
 CRM\ series in Mathematical Physics, Ed. R.\ Conte, Springer Verlag  (1999)

\bibitem{Maciejewski} A. J. Maciejewski,  J.-M. Strelcyn, M. Szydlowski, 
 \emph{J. of Math. Phys.}, \textbf{42},  pp. 1728--1743 (2001)

\bibitem{Uggla}C.\ Uggla,  Chapter 10 of \cite{weell} (1997)

\bibitem{weell}J. Wainwright, G.E.R. Ellis, \emph{Dynamical systems in
cosmology}, Cambridge University Press (1997)

\bibitem{Jantzen}R.T.\ Jantzen,  \emph{Proc. Int. Sch. Phys
E.\ Fermi}, Course LXXXVI (1982), on ''Gamov Cosmology'', R. Ruffini,
F.\ Melchiorri Eds., North Holland, Amsterdam, pp. 61--147, (1987)

\bibitem{ringstrom}H.\ Ringstr\"{o}m,  \emph{Ann. Inst. H. Poincaré}, \textbf{2}, pp. 405  (2001) 

\bibitem{cl}N.J.\ Cornish, J.J.\ Levin, \emph{Phys. Rev. D}, \textbf{55}, pp. 7489  (1997)

\bibitem{bd}J.D. Barrow,\ M.P. Dabrowski, \emph{Phys. Rev. D}, \textbf{57}, pp. 7204  (1998)

\end{thebibliography}
\end{document}